# Lévy flights for light in ordered lasers

Erick G. Rocha,[1] Emanuel P. Santos,[1] Bruno J. dos Santos,[1] Samuel S. de Albuquerque,[1] Pablo I. R. Pincheira,[2] Carlos A. Pereira,[1] and André L. Moura[1,*]

[1]Grupo de Física da Matéria Condensada, Núcleo de Ciências Exatas – NCEx, Campus Arapiraca, Universidade Federal de Alagoas, 57309-005, Arapiraca-AL, Brazil

[2]Departamento de Ciencias Físicas, Universidad de La Frontera, Temuco, Chile

[*]Corresponding author: andre.moura@fis.ufal.br

**Abstract**

Lévy flights for light have been demonstrated in disordered systems with and without optical gain, and remained unobserved in ordered ones. In the present letter, we investigate, numerically and experimentally, Lévy flights for light in ordered systems due to an ordered (conventional) laser. The statistical analysis was performed on the intensity fluctuations of the output spectra upon repeated identical experimental realizations. We found out that the optical gain and the mirrors reflectivity are critical parameters governing the fluctuation statistics. We identified Lévy regimes for gain around the laser threshold, and Gaussian-Lévy-Gaussian crossovers were unveiling when increasing the gain from below to above the threshold. The experimental results were corroborated by Monte Carlo simulations, and the fluctuations were associated to a Langevin noise source that takes into account the randomness of the spontaneous emission, which seeds the laser emission and can cause large fluctuations of the output spectra from shot-to-shot under identical experimental realizations.

Lévy-stable is an interdisciplinary study and has been used to describe problems such as geology, finance, and animal foraging [1]. The central limit theorem states that the sum of random variables with finite variance results in a Gaussian probability distributions. Instead of, the Lévy-stable is explained by the generalized central limit theorem, where the averages may or not be finite, but the variance diverges. The probability to observe a variable $x$ in the limit $x \rightarrow \infty$ is proportional to $x^{-(1+\alpha)}$; Gaussian distributions for $\alpha \geq 2$, and Lévy ones for $0 < \alpha < 2$.

In the photonic context, Lévy flights were reported for light propagating in disordered materials without [2] and with gain [3]. In the former regime, Lévy flights were observed for the transmitted light intensity through a specially designed material where the step-length distribution could be chosen by mixes of high-refractive-index scattering particles ($TiO_2$) and glass microspheres with tailored size distribution [2]. When introducing gain in the disordered materials, the multiple scattering of light provide feedback, and laser action can occur without a resonant cavity, which is called Random Laser [4]. Lévy distributions have been reported for the output intensity of random lasers: under identical experimental realizations the output spectra change stochastically from one realization to another, and the corresponding probability distributions are Gaussian or Lévy [5].

Despites the large quantity of articles concerned with the demonstration and understanding-of Lévy distributions for the output intensity fluctuations of RLs [5-8], to the best of our knowledge, there is not any report in ordered systems. The aim of the present article is to investigate the Lévy distributions for the output intensity fluctuations of ordered (conventional) lasers using as prototype a Q-switched Nd:YAG. The experimental results are supported by Monte Carlo numerical simulations.

The experiments were performed using a Q-switched Nd:YAG laser operating at 1064 nm with pulse duration of ~6 ns. The voltage applied to the flashlamps was fixed at 1550 V with a frequency of 20 Hz. The flashlamps excited the trivalent neodymium ions ($Nd^{3+}$) whose energy level diagram with the excitation and laser transitions are depicted in the Supplemental Material [9]. The optical gain was varied by the delay between the Q-switch (QS) and the flashlamps, that means control the initial density of ions in the upper level of the laser transition ($N_0$). The QS frequency was 10 Hz. Each time the QS was activated, the system was able to lase, and we name it a shot. Keeping the QS delay fixed, each shot is considered an identical realization of the experiment under fairly identical conditions, since the time between the laser shots (10 ms) is much larger than the spontaneous relaxation of the $Nd^{3+}$ that is on the order of 100 µs. The output spectra were acquired using a spectrometer equipped with a charge-coupled-device (CCD) and triggered with the QS. The CCD exposure time was 70 ms which is smaller than the time between optical pulses which ensures single shot spectrum acquisition. The spectral range was from 350 to 1150 nm with resolution of 2.0 nm. For each QS delay, 1000 spectra were acquired to perform the statistical analysis.

The laser characterization was performed by decreasing the QS delay, i.e., increasing the $N_0$ from below the threshold (only spontaneous emission) to well above the threshold (laser regime). Figure 1a shows the average output intensity and the corresponding relative error as a function of the QS delay. The first 30 spectra for each value of the QS delay marked in Fig. 1a are shown in Figs. 1b-d. For Q-switch delay of 364 µs, the gain is below the laser threshold, and one has only spontaneous emission. The spectral intensity fluctuations from shot-to-shot (Fig. 1b) are mild and due to flashlamps intensity fluctuations, which are very small (less than 1% - bottom of Fig. 1a). The corresponding probability distribution for the intensity fluctuations is Gaussian (Fig. 1e). Increasing the gain, by decreasing the QS delay to 356 µs, the gain reaches the threshold for laser emission, and the spectral intensity fluctuation from shot-to-shot is very strong (Fig. 1c) with maximum relative error (Fig. 1a). Noticeably, the corresponding probability distribution is non-Gaussian with a heavy-tail, characteristics of Lévy distributions (Fig. 1f). By increasing the gain further, the intensity fluctuations are smoothed (Fig. 1d for QS delay of 336 µs), the relative error decreases (Fig. 1a), and the shape of the probability distribution changes, recovering the Gaussian profile (Fig. 1g). To analyze the Gaussian-Lévy-Gaussian crossovers in detail, the parameter α, determined by the curve fitting (Eq. 2 of the Supplemental Material [9]), is plotted as a function of the QS delay in Fig. 1a. The Gaussian-Lévy-Gaussian transitions are evident as α changes from 2.0 to 0.4 at the onset of laser emission and from 0.4 back to 2.0 at larger gain.

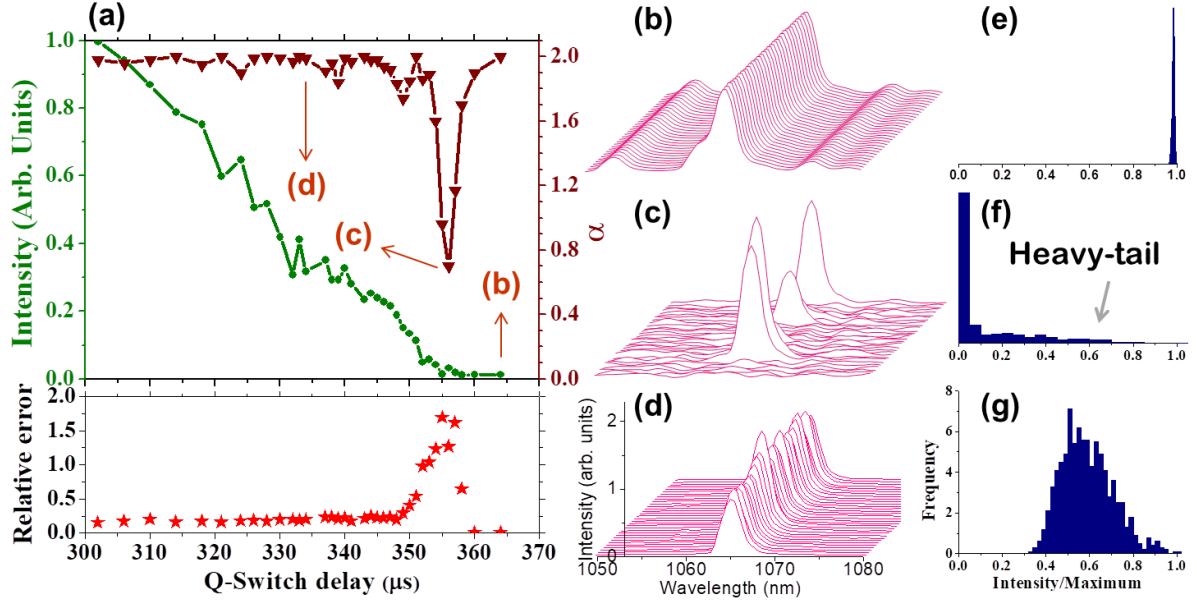

Fig. 1. Characterization of the output of a Nd:YAG laser over 1000 identical realizations for each gain value which was controlled by the delay between the flashlamp and the Q-switch. (a) average output intensity, Lévy exponent (α), and relative error as a function of the Q-switch delay. (b), (c), and (d) shows 30 output spectra for the Q-switch delay values indicated in Fig. 1a. (e), (f), and (g) exhibit the corresponding probability distributions for the output intensity fluctuations considering the 1000 spectra.

The transition from the fluorescence to the laser regime can be described by a simple coupled rate equation system for the population inversion and the energy at the laser transition which is a deterministic system, and the fluctuations are not explained. On the other hand, in Monte Carlo simulations the fluctuations arise naturally. Based on previous outstanding publications on random lasers [6-8], we considered a one dimensional system (1D), and the steps were as follow. (1) The active ions were in the upper level of the laser transition with an initial population density $N_0$ (gain). (2) The density of ions which spontaneously relaxes in a time interval dt is Nτdt, where N is the density of excited ions in the time t, τ is the lifetime of the energy level. We multiplied it by $\beta = 0.1$ to consider that not all of the excited ions contribute to photons at the cavity axis, and to include nonradiative relaxations of the upper energy level of the laser transition. To mimic the linewidth of a real transition, the frequencies of the photons were due to a weighed random Cauchy distribution centered at the resonance one with full width at half maximum of 100 (arbitrary units). (3) The spontaneously created photons whose direction coincides with the cavity axis have initial direction with 50/50 per cent of probability to right/left. Once created, the photons propagate to right or left until the boundary of the 1D medium. (4) Each step of the photons in the amplifying region, they can create other photons by stimulated emission with a rate $\gamma N n_\omega$. (5) At the boundaries, there is probability $\delta$ which varies from 0.0 to 1.0 to the photon return to the gain medium, which can be understood as a measurement of the reflectivity of mirrors. $\delta$ equals 1.0 corresponds to a closed cavity, while for $\delta = 0.0$ there is a completely open system without feedback for laser action. Each time photons scape from the gain medium, the number of photons in each frequency is counted. Each photon evolves independently, and it is essential in the simulations to observe the return from the Lévy to the Gaussian regime. For each pair of values of $N_0$ and $\delta$, one hundred spectra were simulated.

Some of the simulated spectra and the corresponding probability distributions (insets) are shown in Fig. 2. The first, second, and third column correspond to $\delta$ equals 0.0, 0.1, and 0.3, respectively, while the rows correspond to $N_0$ equals $10^6$, $2.5 \times 10^6$, and $5.5 \times 10^6$. Increasing the gain for $\delta = 0.0$ (Figs. 2a-c), we observe only a spectral narrowing and increasing in the output intensity, that is characteristics of amplification of the spontaneous emission during the photons propagation in the gain medium [10]. Noticeably the corresponding probability distributions are Gaussian (insets of Figs. 2a-c). On the other hand, for $\delta = 0.1$ the transition from the fluorescent to the laser regime is evident (Figs. 2d and 2e) by the appearance of narrow peaks with giant intensities (Fig. 2e) and from the crossover in the statistical regimes from Gaussian to a heavy-tail (Lévy) (insets of Figs 2d and 2e). Finally for $\delta = 0.3$ we observe the first transition from a smooth spectrum to a spiky with the change in the probability distributions from Gaussian (Fig. 2g) to Lévy (Fig. 2h), and at gain well-above the threshold, a tendency of spikes suppression with a lot of narrow peaks superimposed to the background of the spontaneous emission and the Gaussian probability distribution (Fig. 2i) in excellent agreement with the experimental results of Fig. 1. The spikes are not observed in the experimental results (Fig. 1b-d) due to the low resolution of the spectrometer. However, the large intensity lasing peaks and very small intensities (non-lasing) under the same experimental conditions observed in Fig. 1c are reproduced (Fig. 2e). In the non-lasing spectra, the energy is distributed among the modes due to intrinsic competition for the available gain, and/or dumped out as non-radiative relaxations and/or as photons whose directions are different from the cavity axis. These two last effects are accounted by the β parameter we considered in the spontaneous emission [11].

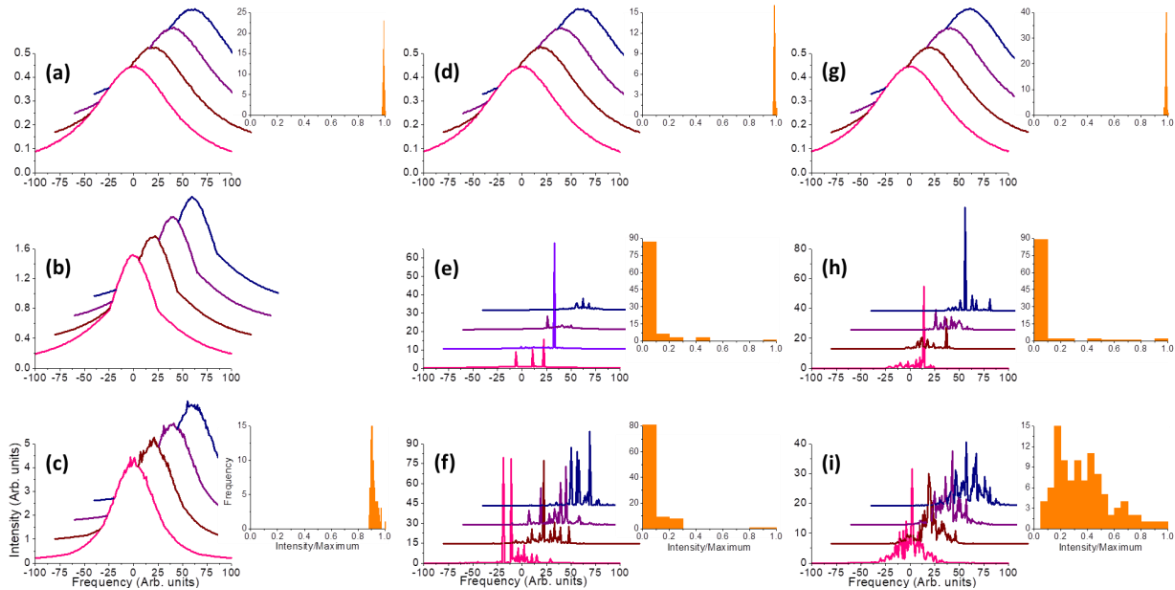

Fig. 2. Output spectra of an ordered laser generated by Monte Carlo simulations. The rows correspond to an initial population $N_0$ equals $10^6$, $2.5 \times 10^6$, and $5.5 \times 10^6$, respectively, and the columns correspond to mirrors reflectivity (δ) of 0.0, 0.1, and 0.3, respectively. The corresponding probability distributions for the intensity fluctuations at the central frequency considering the 100 spectra for each pair of $N_0$ and δ are displayed in the insets.

The influence of the mirrors reflectivity on the output intensity fluctuation of the laser was investigated numerically and is resumed in Fig. 3, which shows the parameter α as a function of the gain ($N_0$) for different values of δ. The transition from spontaneous emission to the laser regime is evident by the cross-over from the Gaussian (α ≥ 2.0) to the Lévy regimes (α < 2.0) (due to the limited number of realizations, we considered Lévy regimes α < 1.8 - dashed line in Fig. 3). This first transition and the return to the Gaussian regime at large gain depend on δ. For δ = 0 the distribution is always Gaussian due to the absence of feedback for laser action. For small values of δ ≠ 0, the return is slow. On the other hand, for δ = 0.5 there is a fast return from the Lévy to the Gaussian regime, and for δ = 0.9 the large fluctuations are suppressed, which agree with recent results on specially designed laser cavities [12].

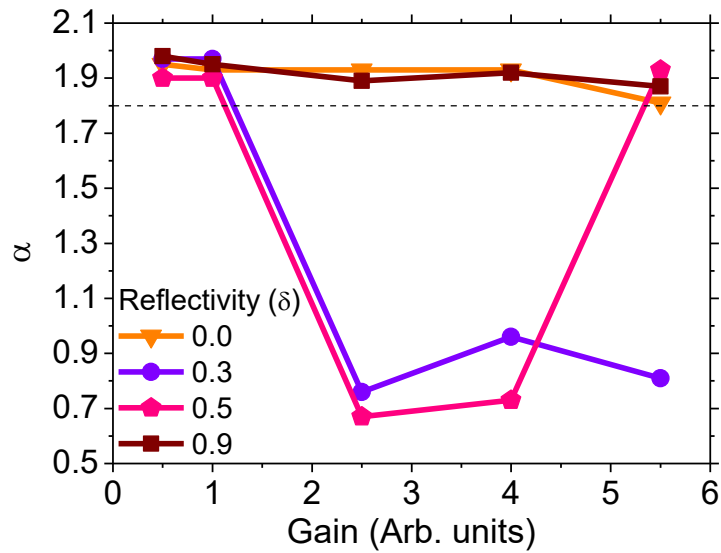

Fig. 3 – Characterization of fluctuation statistics (α ≥ 2.0 – Gaussian, and α < 2.0 – Lévy) for the output intensity of ordered lasers simulated by Monte Carlo method. α parameter is given as a function of the gain ($N_0$) for different values of the mirrors reflectivity (δ). Due to the limited number of simulations for each pair of $N_0$ and δ, we considered the dashed line a guide to decide between Gaussian (α ≥ 1.8) or Lévy (α < 1.8) distributions.

At this point, an analogy with the behavior reported for RLs unveils [8]: the mirrors reflectivity plays a role analogue to the scattering strength in random lasers. In random lasers, the scattering strength has a large influence on observation of Lévy distributions when increasing the gain from below to well above the threshold. For low scattering strength one has the transition from Gaussian to Lévy distributions at the threshold, and a slow return to the Gaussian regime at larger gain. Increasing the scattering strength, there is a fast return to the Gaussian regime, and at much larger scattering strength there is no Gaussian-Lévy crossover when crossing the threshold, i.e. the probability distributions are Gaussian independently for the amount of the optical gain. The return from Lévy to Gaussian regime at large gain is due to the spatial overlapping and temporal coincidence of a large number of optical modes which compete for the available gain, depleting it and suppressing rare strong amplifications.

Summing up, in the ordered lasers (lasers with cavities), one has that for gain just above the threshold, just few optical modes are activated, and the rare large peak intensities are due to "lucky" photons [13] with random frequency which take the gain and dominate the emitted spectrum. In each realization of the experiment under fairly identical conditions, the frequency of the "lucky" photon is a random variable weighted by the spontaneous emission curve (Fig. 2e). Increasing the gain from below to above the threshold occurs the first statistical distribution transition from Gaussian to Lévy. By increasing the gain further, there is a large spatial overlapping among the optical modes, which depletes the gain and leads to self-averaged Gaussian distributions. For $\delta$ closes to 1.0, the fraction of spontaneously emitted photons which are feed back to the gain volume is large, implying in a large spatial overlapping of the optical modes even for gain in the vicinity of the threshold. Then, the probability distributions are Gaussian in all regimes, i.e. fluorescent and lasing ($\alpha \approx 2.0$ for $\delta = 0.9$ in Fig. 3). In this context the intensity fluctuations can be understand by a *Langevin noise source* due to randomness of the spontaneous emission.

It is worth emphasizing that non-Gaussian distributions were reported for conventional lasers but the underlying mechanism was due to random variation of the gain from one experimental realization to another [14], which can be attributed for example to random fluctuations of the excitation source. Recently, intensity fluctuations in ordered lasers were deeply investigated in Refs. [10,15] aiming at the demonstration of replica symmetry breaking. The strong fluctuations were associated to the nonlinear interaction among the optical modes by the third-order nonlinear susceptibility of the gain medium. In the present letter, the fluctuations were described without take into account the phases of the modes, i.e., only intensity feedback was considered.

In summary, Lévy flights for light in ordered systems were demonstrated for the first time due to cavity (conventional) lasers. Lévy flights are due to the stochasticity of the spontaneous emission. Lévy distributions were unveiled when the fraction of emitted photons fed back to the gain medium was small, and for gain around the threshold. Under these conditions, the fraction of spontaneously emitted photons which is feed back into the gain volume is small implying in fluctuations in the observable quantities. The present results show the Lévy distributions are not exclusive of disordered systems, and can enlightening the discussion of optical mode competition for the available gain with nonlinear interaction among them and frustration induced by disorder in the photonic-magnetic analogy [16]. Finally, the present results open several possibilities for the connection between lasers and the statistical physics of complex systems, and signalize that studies until now restricted for disordered (random) lasers can be performed in closed cavity ones, like the recent demonstrated turbulence hierarchy in random lasers [17].

We acknowledge financial support from the Brazilian Agencies: Conselho Nacional de Desenvolvimento Científico e Tecnológico (CNPq), and Fundação de Amparo à Pesquisa do Estado de Alagoas (FAPEAL). We thank to J. G. B. Cavalcante, M. A. Nascimento, and people from Quantumtech (especially J. Rotava and D. Costa) for technical support. A. L. M. is very grateful to C. B. de Araújo, A. S. L. Gomes, and E. P. Raposo. P. I. R. Pincheira acknowledges support from FONDECYT No. 3180696.